\documentclass[conference]{IEEEtran}
\IEEEoverridecommandlockouts
\usepackage{cite}
\usepackage{amsmath,amssymb,amsfonts}
\usepackage{algorithmic}
\usepackage{graphicx}
\usepackage{textcomp}
\usepackage{xcolor}
\usepackage{amsthm}
\usepackage{multirow}
\usepackage{diagbox}
\usepackage{colortbl}
\usepackage{booktabs}
\usepackage{arydshln}
\usepackage[colorlinks=true,linkcolor=black]{hyperref}

\usepackage{float}
\usepackage{subcaption}

\def\BibTeX{{\rm B\kern-.05em{\sc i\kern-.025em b}\kern-.08em
    T\kern-.1667em\lower.7ex\hbox{E}\kern-.125emX}}
\begin{document}

\title{Enhancing Deep Learning Performance of Massive MIMO CSI Feedback}

\author{\IEEEauthorblockN{Sijie Ji}
\IEEEauthorblockA{
\textit{Nanyang Technological University}\\
Singapore \\
sijie001@e.ntu.edu.sg}
\and
\IEEEauthorblockN{Mo Li}
\IEEEauthorblockA{
\textit{Nanyang Technological University}\\
Singapore \\
limo@ntu.edu.sg}}

\maketitle

\begin{abstract}

CSI feedback is an important problem of massive multiple-input multiple-output (MIMO) technology because the feedback overhead is proportional to the number of sub-channels and the number of antennas, both of which scale with the size of the massive MIMO system.Deep learning-based CSI feedback methods have been widely adopted recently owing to their superior performance.
Despite the success, current approaches have not fully exploited the relationship between the characteristics of CSI data and the deep learning framework.

In this paper, we propose a jigsaw puzzles aided training strategy (JPTS) to enhance the deep learning-based massive MIMO CSI feedback approaches by maximizing mutual information between the original CSI and the compressed CSI.
We apply JPTS on top of existing state-of-the-art methods. Experimental results show that by adopting this training strategy, the accuracy can be boosted by 12.07\% and 7.01\% on average in indoor and outdoor environments, respectively. The proposed method is ready to adopt to any existing deep learning frameworks of massive MIMO CSI feedback.
Codes of JPTS are available on GitHub for use\footnote{\url{https://github.com/SIJIEJI/JPTS}}.

\end{abstract}


\IEEEpeerreviewmaketitle

\section{Introduction}
\IEEEPARstart{T}he massive multiple-input multiple-output (MIMO) is one of the key technologies for next generation communication systems, e.g., 5G and above. Unlike traditional cell-based communication paradigms, the massive MIMO makes better use of spatial diversity and serves users in a cell-free way.
A massive MIMO system typically is equipped with a large number of antennas at the base station (BS), which aims to make full use of spatial diversity by conducting beamforming to concentrate signal energy to a specific user equipment (UE).

BS requires the downlink channel state information (CSI) to conduct beamforming. Especially for modern cellular systems that work on frequency division duplexing (FDD) mode, the UE may have to explicitly feed back the downlink CSI to the BS due to the lack of channel reciprocity.
As the overhead of CSI feedback grows quadratically with the number of transmitting antennas, CSI compression is needed before the feedback to reduce the overhead.
Therefore, how to feedback CSI efficiently and accurately becomes one most important problem for massive MIMO~\cite{lu2014overview}. 

\begin{figure}[h]
  \centering
\includegraphics[scale=0.55]{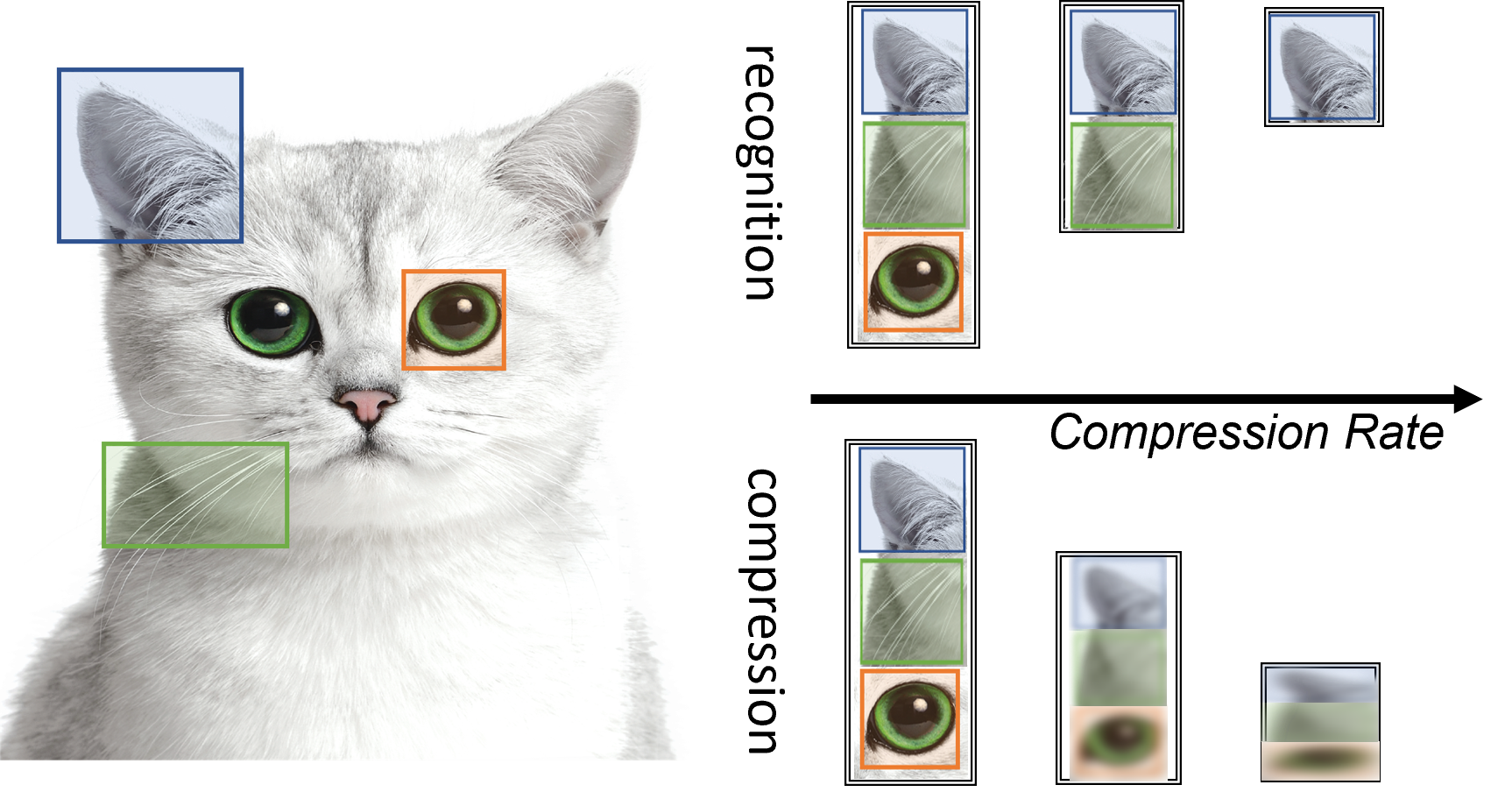}
  \caption{The Difference between recognition task and compression task. With the increased compression ratio, a recognition task aims to keep the most distinguishable features of the original data while the compression task aims to keep as complete (but approximated) information as possible.}
  \label{f:motiv}
\end{figure}

Traditional compressive sensing (CS) based CSI feedback methods rely on specific assumption and require iterative reconstruction of CSI, which makes it hard to adapt to fast-changing channels in real life. 
Ever since the first DL-based method CSINet~\cite{wen2018deep} was proposed, a series of subsequent works~\cite{ji2021clnet,lu2020multi,cui2022transnet,bi2022novel} appear because of the superior performance brought by deep learning. Generally, DL-based methods utilize the auto-encoder framework~\cite{hinton2006reducing}, where the encoder learns to compress the original CSI at the UE side and the decoder learns to reconstruct the original CSI at the BS side. The auto-encoder is trained in unsupervised manner without the need for labeled data and only requires a single run upon deployment for continuous CSI reconstruction, which overcomes the computation inefficiency of traditional CS-based approaches.

Despite the success of DL-based CSI feedback methods, current approaches have not fully exploited the advantages of deep learning due to the lack of analysis of the CSI feedback problem and the characteristics of CSI data. 
Current DL-based approaches generally migrated some of the well-known neural network building blocks, such as residual blocks~\cite{wen2018deep}, transformer~\cite{bi2022novel}, LSTM~\cite{wang2018deep} and \textit{etc.}, to the CSI feedback task.
Those modules are proposed mainly for the recognition problem in computer vision or natural language processing domain, which differs from the CSI compression task. As Fig.~\ref{f:motiv} illustrates, the recognition task aims to capture the most distinctive features of the original data  while the compression task should aim at preserving the information approximately but as complete as possible. 

Especially for the CSI data that is usually sparse, the encoder tends to discard sparse parts when doing compression, then zero-pads those parts during reconstruction. Although the absolute value of the original CSI matrix and the reconstructed CSI matrix are similar by doing the zero-pad operation, the reconstruction error depends on the relative position of those zero-padded parts. In addition, the CSI matrix carries physical information with the delays and angles of the propagation paths, so the relative position of sparse parts matters. As a result, the physical information carried by CSI becomes inaccurate when the encoder discards more parts without the knowledge of the relative positions of them in the CSI matrix.
Therefore, improving performance of CSI feedback task requires a mechanism that helps retain complete but approximative information rather than discarding relatively unimportant information and only preserving the most distinguishable parts.
In particular, the physical information of path delays and angles that are encoded as position information in the CSI matrix should be considered.

To this end, we propose a jigsaw puzzles aided training strategy (JPTS), which involves an auxiliary jigsaw puzzle-solving task during training so the neural network is forced to fuse the position information across different local regions (puzzle pieces) of the CSI matrix while encoding the most representative information so that even the model has better knowledge about where to zero-pads when reconstruction and preserve the holistic CSI information as complete (but approximated) as possible.
We evaluate the effectiveness of JPTS by adopting it on top of three open sourced SOTA CSI feedback methods and experiment with both the indoor and outdoor CSI data. The experimental results demonstrate that the proposed JPTS effectively improves the performance of SOTA CSI feedback approaches. The strategy helps to improve the overall average accuracy of CSINet, CRNet and CLNet by 25.66\%, 6.98\% and 3.58\% for indoor scenario and 16.67\%, 2.18\% and 2.17\% for outdoor scenario, respectively. There is an overall accuracy lift of 14.80\%, 6.32\%, 10.14\%,   10.09\% and   6.35\%  at corresponding compression ratios of 1/4, 1/8, 1/16, 1/32 and 1/64, respectively. The highest improvement is 39.43\% which is obtained from outdoor JPTS-CSINet with 1/4 compression ratio.


The main contributions are summarized as follows:
\begin{itemize}
\item To the best of our knowledge, JPTS is the first training strategy proposed for massive MIMO CSI feedback.
\item The proposed JPTS helps improve the performance of DL-based CSI feedback approaches.
\item JPTS is effective to different deep neural network architectures and thus can be generally applied across different SOTA approaches for DL-based CSI feedback.

\end{itemize}

\section{System Model}

For simplicity, a single cell massive MIMO system operating in FDD mode is considered, where the BS is equipped with $N_{t}$ antennas and the UE side has $N_{r}$ antennas. $N_{t}$ $\gg N_{r}$ ($N_{r}$ equals to 1 for simplicity). The orthogonal frequency division multiplexing (OFDM) is adopted with $N_{c}$ subcarriers. The received signal $\textbf{\textit{y}} \in \mathbb{C}^{ N_{c} \times 1}$ can be expressed as follows:
\begin{equation}
\textbf{\textit{y}}=\textit{\textbf{A}} \textbf{\textit{x}}+\textbf{\textit{z}}
\label{e:received_signal}
\end{equation}
where $\textbf{\textit{x}} \in \mathbb{C}^{N_{c} \times 1}$ indicates the transmitted symbols and $\textbf{\textit{z}} \in \mathbb{C}^{N_{c} \times 1}$ is the complex additive Gaussian noise. 
$\textit{\textbf{A}}$ is a diagonal matrix that can be expressed as $\operatorname{diag}\left(\textbf{\textit{h}}_{i}^{H} \textbf{\textit{p}}_{i}, \cdots, \textbf{\textit{h}}_{N_{c}}^{H} \textbf{\textit{p}}_{N_{c}}\right)$,  $i \in\left\{1, \cdots, N_{c}\right\}$,
where $\textbf{\textit{h}}_{i} \in \mathbb{C}^{N_{t} \times 1}$ is the downlink channel coefficients and $\textbf{\textit{p}}_{i} \in \mathbb{C}^{N_{t} \times 1}$ represent beamforming precoding vector for subcarrier $i$. 

Due to the asymmetry of uplink and downlink, in order to derive the beamforming precoding vector $\textbf{\textit{p}}_{i}$, the BS needs the knowledge of corresponding downlink channel coefficient $\textbf{\textit{h}}_{i}$ that is fed back by the UE.
The downlink channel matrix is $\textit{\textbf{H}}=\left[\textbf{\textit{h}}_{1} \cdots \textbf{\textit{h}}_{N_{c}}\right]^{H}$ which contains $N_{c}N_{t}$ elements. As the CSI matirx $\textit{\textbf{H}}$ is complex-valued, the total number of parameters that need to be fed back is $2N_{c}N_{t}$, which is proportional to the number of antennas.
Since the characteristic of massive MIMO is assembled with an extremely large antenna array, the overhead of direct CSI feedback is unacceptable, and thus how to better compress CSI becomes the bottleneck problem to enable massive MIMO.

The common practice is first to obtain angular-delay domain CSI representation $\textit{\textbf{H}}^{\prime}$, because the channel matrix $\textit{\textbf{H}}$ is often sparse in the angular-delay domain. $\textit{\textbf{H}}^{\prime}$ can be obtained by performing the 2D discrete Fourier transform (DFT) on $\textit{\textbf{H}}$ such that
\begin{equation}
\textit{\textbf{H}}^{\prime}=\textit{\textbf{F}}_{c} \textit{\textbf{H}} \textit{\textbf{F}}_{t}^{H}
\end{equation}
where $\textit{\textbf{F}}_{c} \in \mathbb{C}^{N_{c} \times N_{c}}$ and $\textit{\textbf{F}}_{t}  \in \mathbb{C}^{N_{t} \times N_{t}}$ are the DFT transform matrices. Each element in $\textit{\textbf{H}}^{\prime}$ represents a certain path delay with a certain angle of arrival (AoA). Since the time delay of multi-path arrivals is within a finite time, only the first few rows contain useful information, while the rest rows are made up of near-zero values, and can be omitted without much information loss. We may obtain the informative $\textit{\textbf{H}}^{\prime}$ as $\textit{\textbf{H}}_{a} \in \mathbb{C}^{N_{t} \times N_{t}}$. However, even after that, $\textit{\textbf{H}}_{a}$ is still too big to feedback because $N_{t}$ is large.

The DL-based solution therefore achieves affordable overhead by assembling an encoder on the UE side for $\textit{\textbf{H}}_{a}$ compression and a decoder on the BS side for $\textit{\textbf{H}}_{a}$ reconstruction. Mathematically, 
\begin{equation}
\hat{\textit{\textbf{H}}}_{a}=f_\mathcal{D}\left(f_\mathcal{E}\left(\textit{\textbf{H}}_{a}, \mathit{\Theta}_{\mathcal{E}}\right), \mathit{\Theta}_{\mathcal{D}}\right)
\end{equation}
where $f_{\mathcal{E}}$ and $f_{\mathcal{D}}$ denote the encoding process and the decoding process, respectively, and $\mathit{\Theta}_{\mathcal{E}}$ and $\mathit{\Theta}_{\mathcal{D}}$ represent a set of learned parameters of the encoder and the decoder, respectively. The output of the encoder is the compressed CSI that is denoted as $\textit{\textbf{v}}$. 
The compression ratio $\eta$ is defined as the ratio of $\textit{\textbf{v}}$ and the encoder's input such that:
\begin{equation}
\eta=\frac{\textit{\textbf{v}}}{2 N_{t} N_{t}}
\end{equation}

\begin{figure*}[h!]
  \centering
\includegraphics[scale=0.48]{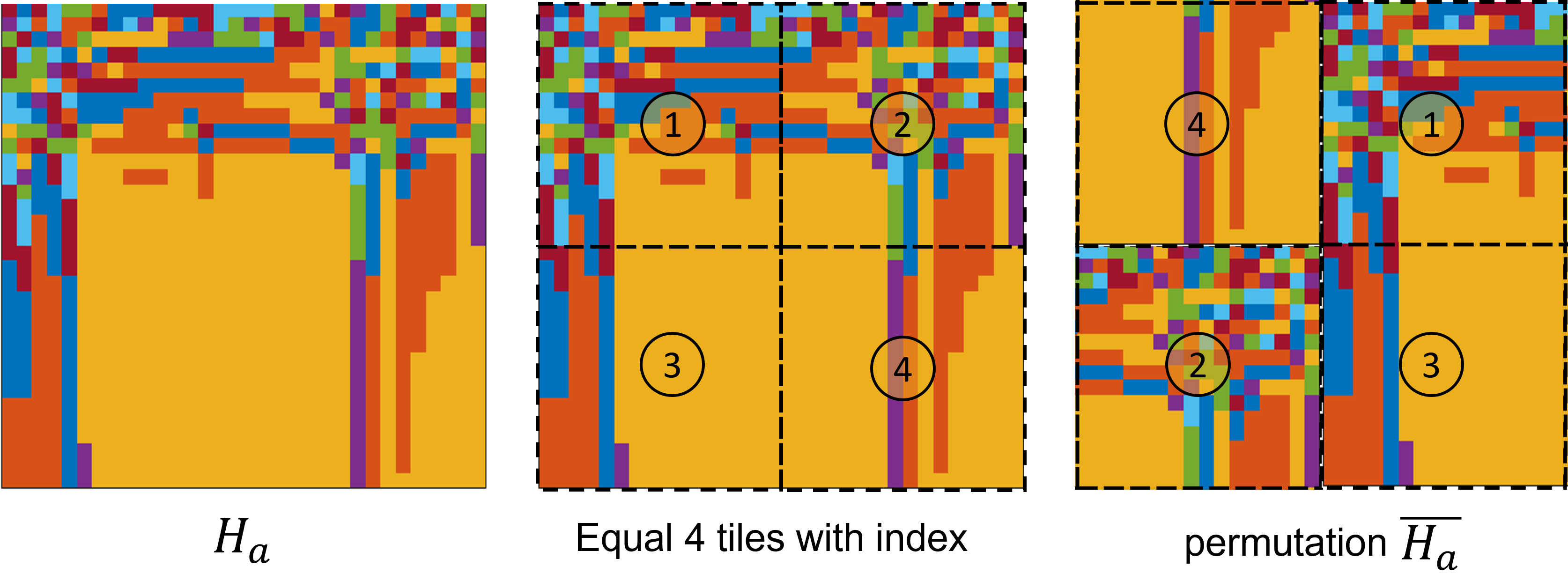}
  \caption{Jigsaw Puzzles aided Training Strategy: The original $\textit{\textbf{H}}_{a}$ is first divided into n tiles ($n=4$ as an example) with equal size, and the 4 tiles are randomly shuffled to derive $\overline{\textit{\textbf{{H}}}_{a}}$, the permutation version of $\textit{\textbf{H}}_{a}$. The auxiliary task requires the encoder to provide the index of the permutation to solve the jigsaw puzzles correctly so that the encoder should gain the relative position knowledge when compressing the original $\textit{\textbf{H}}_{a}$}
  \label{f:jigsaw}
\end{figure*}
The goal of DL-based solutions is to find the parameter sets of encoder and decoder with minimize the difference between the original $\textit{\textbf{H}}_{a}$ and the reconstructed $\hat{\textit{\textbf{H}}}_{a}$:
\begin{equation}
\left(\hat{\mathit{\Theta}}_{\mathcal{E}}, \hat{\mathit{\Theta}}_{\mathcal{D}}\right)=\underset{\mathit{\Theta}_{\mathcal{E}}, \mathit{\Theta}_{\mathcal{D}}}{\arg \min }\left\|\textit{\textbf{H}}_{a}-f_{\mathcal{D}}\left(f_{\mathcal{E}}\left(\textit{\textbf{H}}_{a}, \mathit{\Theta}_{\mathcal{E}}\right), \mathit{\Theta}_{\mathcal{D}}\right)\right\|_{2}^{2}
\end{equation}

\section{Jiwsaw Puzzles Training Strategy}
The compression task essentially is to maximize the mutual information (MI) between the original input and the compressed output~\cite{hjelm2018learning}, which differs from the recognition task that focus to learn the most distinctive part of the input. The MI between the input $\textit{\textbf{H}}_{a}$ and output $\textit{\textbf{v}}$ of the encoder is defined as: 
\begin{equation}
I(\textit{\textbf{H}}_{a} ; \textit{\textbf{v}})=\sum_{\textit{\textbf{H}}_{a}, \textit{\textbf{v}}} p(\textit{\textbf{H}}_{a}, \textit{\textbf{v}}) \log \frac{p(\textit{\textbf{H}}_{a}, \textit{\textbf{v}})}{p(\textit{\textbf{H}}_{a}) p(\textit{\textbf{v}})}
\end{equation}
which measures the inherent dependence of the joint distribution of $\textit{\textbf{H}}_{a}$ and $\textit{\textbf{v}}$ relative to the marginal distribution of $\textit{\textbf{H}}_{a}$ and $\textit{\textbf{v}}$ under the assumption of independence. $I(\textit{\textbf{H}}_{a} ; \textit{\textbf{v}})=0$ means $\textit{\textbf{H}}_{a}$ and $\textit{\textbf{v}}$ are independent. 
Base on the definition, larger MI indicates less uncertainty between $\textit{\textbf{H}}_{a}$ and $\textit{\textbf{v}}$, higher amount of information obtained of  $\textit{\textbf{v}}$  through observing the $\textit{\textbf{H}}_{a}$.

The current DL-based CSI compression methods use auto-encoder structure with reconstruction loss, which in fact maximizes the MI in a global way~\cite{shwartz2017opening}. Later research demonstrates that maximizing the MI in local regions (e.g. patches rather than the complete
image) can greatly improve the representation’s quality~\cite{noroozi2016unsupervised}, which also explains why various attention mechanisms~\cite{ji2021clnet,cui2022transnet} and multi-resolution~\cite{lu2020multi} blocks are useful in CSI feedback task. Nevertheless, the attention mechanisms only focu on the local patche/cluster with signal path information while those sparse parts are simply ignored. Although the value of sparse parts is negligible, the position information of sparse parts matters because they actually help determine the position of signal paths that carry the physical information of path delays and angles. 

Therefore, we introduce the idea of jigsaw puzzles, which were first introduced by John Spilsbury as a pretext for learning geography. Solving jigsaw puzzles is another way to maximize MI in a local way with the natural benefit of learning position information. 
In order to solve the puzzles correctly, the network should have the ability to recognize the relative positions of different parts of the original data, whose characteristic is suitable for CSI feedback task as the positions of CSI sub-matrices also carry physical information. Currently, using deep learning to solve jigsaw puzzles by predicting the location of puzzle fragments is frequently used as a self-supervised learning pretext for representative learning~\cite{he2020momentum}. 
We design a jigsaw puzzle-solving task as an auxiliary task in the training phase to enforce the network to encode the relative position even for those sparse parts, namely, Jigsaw Puzzles aided Training Strategy for the CSI feedback task.

Concretely, the original CSI matrix $\textit{\textbf{H}}_{a} \in \mathbb{C}^{N_{t} \times N_{t}}$ is divided to $n$ equal-size tiles $T$ followed by a random shuffle to get the permutation index. The learning goal is to be able to get the original version from the shuffled version. The $n$ determine the difficulty of the solving problem, if the number of tiles is 9, there are 9!=362,880 possible permutations. In particular, as Fig.~\ref{f:jigsaw} illustrates, we use $n=4$ as an example. First, the input $\textit{\textbf{H}}_{a}$\footnote{This is the CSI data from the dataset visualized with matlab's 'lines' colormap.} is  divided into four equal-size tiles $\textit{\textbf{T}}_{i} \in \mathbb{R}^{16\times16}$, $i=1,...,4$, followed by a random shuffle, \textit{e.g.} $\textbf{\textit{s}} = [4,1,2,3]$, to rearrange the position of each tile to get the permutation version of $\textit{\textbf{H}}_{a}$ termed as $\overline{\textit{\textbf{{H}}}_{a}}$.
Finally, $\textit{\textbf{H}}_{a}$ and $\overline{\textit{\textbf{{H}}}_{a}}$ are paired together to form the input of the network for training. The auxiliary jigsaw puzzle-solving task lets the encoder recognize the original position of each tile by giving a list of integer index $\textbf{\textit{l}}$ that is the same as the actual permutation index $\textbf{\textit{s}}$. Concretely, we add an extra fully connected layer, $f_{S}$, at the end of the encoder to structure the output to be a matrix with a fixed dimension $\textbf{\textit{J}} \in \mathbb{R}^{4 \times4}$. The matrix further goes through the Softmax function $\sigma$ to re-scale each column of the matrix so they lie in the range of [0,1] and sum to 1. In the end, each column of the matrix is a vector with the probability value of the position for each tile.

The goal of the training now turns to finding the parameter sets of encoder and decoder that not only minimize the difference between the original $\textit{\textbf{H}}_{a}$ and the reconstructed $\hat{\textit{\textbf{H}}}_{a}$ but also minimize the error of the permutation prediction such that:
\begin{equation}
\begin{aligned}
\label{e:all}
\footnotesize
\left(\hat{\mathit{\Theta}}_{\mathcal{E}}, \hat{\mathit{\Theta}}_{\mathcal{D}}, \hat{\mathit{\Theta}}_{\mathcal{S}}\right)=\underset{\mathit{\Theta}_{\mathcal{E}}, \mathit{\Theta}_{\mathcal{D}},\mathit{\Theta}_{\mathcal{S}}}{\arg \min }
[ \alpha \left\|\textit{\textbf{H}}_{a}-f_{\mathcal{D}}\left(f_{\mathcal{E}}\left(\textit{\textbf{H}}_{a}, \mathit{\Theta}_{\mathcal{E}}\right), \mathit{\Theta}_{\mathcal{D}}\right)\right\|_{2}^{2} \\
+ (1-\alpha) \sum_{i=1}^{4}\left\|\textbf{\textit{s}}_{i} -argmax(\sigma(\textbf{\textit{J}}(:,i)))\right\|_{2}^{2}]
\end{aligned}
\end{equation}
in which,
$\textbf{\textit{J}}=f_{S}\left(f_{\mathcal{E}}\left(\overline{\textit{\textbf{{H}}}_{a}}, \mathit{\Theta}_{\mathcal{E}}\right),\mathit{\Theta}_{\mathcal{S}}\right)$ and the $argmax$ denotes getting the index of the maximum element in the vector. The $\alpha$ in  equation~\ref{e:all} is a weighting parameter that controls the difficulty of the jigsaw puzzle-solving task. If $\alpha \to 0$, the training degrades to learning the 2D absolute position of the tiles without learning their semantic content. If $\alpha \to 1$, the training degrades to the original task where the encoder just keeps more discriminative parts and simply discards those sparse parts.

As such, in order to meet the goal of reconstructing $\textit{\textbf{H}}_{a}$ and solving the puzzle correctly at the same time, which requires the encoder to preserve the local region information for recognizing each tile and its relative position within the CSI matrix, instead of just preserving the most discriminative parts of the entire CSI matrix. Especially for CSI data, most part of the $\textit{\textbf{H}}_{a}$ is sparse, e.g. yellow annotated part in Fig.~\ref{f:jigsaw}, without the JPTS, the encoder may simply discard the sparse part.
The decoder just zero-pads the discarded sparse region during reconstruction as the absolute value almost the same. 
By adding the auxiliary jigsaw puzzle-solving task, the sparse part may also be discarded, but the relative position information is preserved by the encoder so that the decoder has better knowledge of where to zero-pad those sparse parts and as a result benefit to the entire CSI matrix. The JPTS maximizes the MI in local regions, which is complementary to the auto-encoder framework that maximizes the MI in a global way. As a result, the whole MI is maximized and the power of those deep learning building blocks is unleashed.


\begin{table*}[ht!]
\centering

\normalsize
\begin{tabular}{crrrrrrrrrr} 
\hline\hline
$\eta$                   & \multicolumn{2}{c}{1/4}                                  & \multicolumn{2}{c}{1/8}                                  & \multicolumn{2}{c}{1/16}                                 & \multicolumn{2}{c}{1/32}                                 & \multicolumn{2}{c}{1/64}                                  \\ 
\hline\hline
\multirow{2}{*}{Methods} & \multicolumn{2}{c}{NMSE}                                 & \multicolumn{2}{c}{NMSE}                                 & \multicolumn{2}{c}{NMSE}                                 & \multicolumn{2}{c}{NMSE}                                 & \multicolumn{2}{c}{NMSE}                                  \\
                         & \multicolumn{1}{c}{indoor} & \multicolumn{1}{c}{outdoor} & \multicolumn{1}{c}{indoor} & \multicolumn{1}{c}{outdoor} & \multicolumn{1}{c}{indoor} & \multicolumn{1}{c}{outdoor} & \multicolumn{1}{c}{indoor} & \multicolumn{1}{c}{outdoor} & \multicolumn{1}{c}{indoor} & \multicolumn{1}{c}{outdoor}  \\ 
\hline\hline
CSINet~\cite{wen2018deep}                   & -17.36                     & -8.75                       & -12.70                     & -7.61                       & -8.65                      & -4.51                       & -6.24                      & -2.81                       & -5.84                      & -1.93                        \\
JPTS-CSINet              & -24.19                     & -12.20                      & -15.20                     & -7.97                       & -10.65                     & -5.22                       & -8.59                      & -3.12                       & -6.26                      & -2.17                        \\ 
\hline
CRNet~\cite{lu2020multi}                    & -24.10                     & -12.57                      & -15.04                     & -7.94                       & -10.52                     & -5.36                       & -8.90                      & -3.16                       & -6.23                      & -2.19                        \\
JPTS-CRNet               & -26.84                     & -12.72                      & -16.32                     & -8.01                       & -11.55                     & -5.41                       & -8.98                      & -3.38                       & -6.50                      & -2.21                        \\ 
\hline
CLNet~\cite{ji2021clnet}                    & -29.16                     & -12.88                      & -15.60                     & -8.29                       & -11.15                     & -5.56                       & -8.95                      & -3.49                       & -6.34                      & -2.19                        \\
JPTS-CLNet               & -28.38                          & -12.90                      & -16.03                     & -8.40                       & -12.16                     & -5.61                       & -9.00                      & \textbf{-3.61}                       & \textbf{-6.86}                      & \textbf{-2.30}                        \\ 
\hline
CSIFormer~\cite{bi2022novel}                & \multicolumn{1}{c}{/}      & \multicolumn{1}{c}{/}       & \multicolumn{1}{c}{/}      & \multicolumn{1}{c}{/}       & \multicolumn{1}{c}{/}      & \multicolumn{1}{c}{/}       & -9.23                      & -3.51                       & -6.85                      & -2.25                        \\
\hline

\end{tabular}

\caption[Caption for LOF]{NMSE(dB) comparison between SOTA CSI feedback approaches and their JPTS enhanced result. / means the performance is not reported.}
\label{t:all}
\end{table*}

\section{Evaluation}
This section details the experiment setting and reports the results of utilizing Jigsaw Puzzles aided Training Strategy with existing open-source state-of-the-art (SOTA) DL-based CSI feedback approaches.

\subsection{\textbf{Data and Evaluation Metric}} 

We conduct experiments based on the most famous open-source dataset in the massive MIMO CSI feedback domain. The dataset is generated based on the COST 2100 channel model~\cite{liu2012cost} and is published with the first DL-based work in this domain named CSINet~\cite{wen2018deep}.  The transmission antennas in the BS side are configured as $N_{t}$ = 32 uniform linear array (ULA) and the receiving antenna $N_{r}$ = 1 in the UE side and the sub-carriers $N_{c}$ = 1024. 
There are two types of scenarios,  indoor pico-cell scenario operating on 5.3~GHz band and outdoor rural scenario operating on 300~MHz band, respectively.
The generated CSI matrices are converted to angular-delay domain $\textit{\textbf{H}}_{a} \in \mathbb{R}^{32 \times 32 \times 2}$ by 2D-DFT. 
The total 150,000 independently generated CSI are split into three parts, i.e., 100,000 for training, 30,000 for validation, and 20,000 for testing, respectively.
The metric to evaluate the performance of CSI reconstruction is the normalized mean square error ($\mathrm{NMSE}$) between the original $\textit{\textbf{H}}_{a}$ and the reconstructed $\hat{ \textit{\textbf{H}}}_{a} $ such that:
\begin{equation}
    \mathrm{NMSE}=\mathrm{E}\left\{\|\textit{\textbf{H}}_{a}-\hat{\textit{\textbf{H}}}_{a}\|_{2}^{2} /\|\textit{\textbf{H}}_{a}\|_{2}^{2}\right\}
\end{equation}
%
We evaluate the efficacy of JPTS by adding the training strategy on top of 3 open-source SOTA DL-based methods, CSINet, CRNet and CLNet, respectively. All models were trained with the batch size of 200 and epoch of 1000 (the same as the three original work) on a single NVIDIA 2080Ti GPU. 
\subsection{\textbf{Overall Performance of JPTS}}

We verify the proposed  jigsaw puzzles aided training strategy by adopting it on top of three SOTA approaches with different compression ratios under both indoor and outdoor scenarios. Table I reports the results. As shown in the Table, in general, JPTS helps improve the overall average accuracy of CSINet~\cite{wen2018deep}, CRNet~\cite{lu2020multi} and CLNet~\cite{ji2021clnet} methods by 25.66\%, 6.98\% and 3.58\% for indoor scenarios and 16.67\%, 2.18\% and 2.17\% for outdoor scenarios. In the indoor scenario, SOTA approaches with the help of JPTS are improved by  16.01\%, 10.32\%, 14.42\%,    13.04\%  and 6.58\% on average, corresponding to 1/4, 1/8, 1/16, 1/32 and 1/64 compression ratio respectively.
In outdoor scenario, they are improved by 13.59\%, 2.31\%, 5.86\% ,  7.14\%  and 6.12\% on average, corresponding to 1/4, 1/8, 1/16, 1/32 and 1/64 compression ratio respectively.
The highest improvement is achieved with indoor JPTS-CSINet and outdoor JPTS-CSINet respectively. Compared to the original scheme, the performance is improved by 39.34\% and 39.43\% at 1/4 compression ratio.

The three SOTA approaches are proposed successively. The CRNet proposes multi-resolution blocks to improve the CSINet and the CLNet proposes attention mechanisms to improve CRNet. As a comparison, by simply adopting the proposed jigsaw puzzles aided training strategy without changing the network structure, the JPTS-CSINet outperforms the CRNet with all compression ratios, which validates our argument. Although multi-resolution is a way to maximize the MI in a local way, solving jigsaw puzzles is a better alternative to maximize the MI in terms of the CSI feedback task. The JPTS-CRNet surpasses CLNet under most compression ratios except for $\eta = 1/4$ indoors also validates the argument and suggests that solving jigsaw puzzles surpass the attention mechanism, even though the CLNet utilizes two different attention mechanisms at the same time. 
At the same time, it is worth noting that the JPTS-CSINet has been improved much more than the JPTS-CRNet and JPTS-CLNet about 6~7 times. This is because JPTS essentially helps maximize the MI in a local way, where CRNet uses multi-resolution blocks and CLNet uses attention mechanisms to do so, but CSINet does not apply any techniques to achieve it, instead, it only able to maximize the MI in a global way. 

Besides, the JPTS-CSINet achieves -6.26 NMSE at the compression ratio of 1/64, where the CSINet achieves a similar NMSE -6.24 at the compression ratio of 1/32. Meanwhile, the NMSE of JPTS-CSINet with 1/32 compression ratio is better than the NMSE of CSINet with 1/16 compression ratio. These indicate that the JPTS helps save the precious network resources and achieve higher compression ratio with similar accuracy.

The most recent approach, CSIFormer~\cite{bi2022novel}, migrates the recent advanced transformer building blocks~\cite{vaswani2017attention} to specifically improve the CSI feedback performance at the high compression ratios (1/32 and 1/64). Similar to the JPTS, the transformer helps to fuse the position information, and thus the accuracy under high compression ratios is better than CLNet with the cost of doubling the complexity of CLNet. As reported in Table~\ref{t:all}, by changing the training method of CLNet, the performance of JPTS-CLNet at 1/64 compression ratio outperforms the CSIFormer by 2.85\% and 2.22\% indoors and outdoors, respectively. The JPTS-CLNet also outperforms the CSIFormer at 1/32 compression ratio when outdoors. The result suggests that the relative position matters, JPTS is a more desired way to learn relative position information without increasing the model complexity. 

\begin{figure}[t]
    \begin{subfigure}[b]{0.49\linewidth}
		\includegraphics[width=\linewidth]{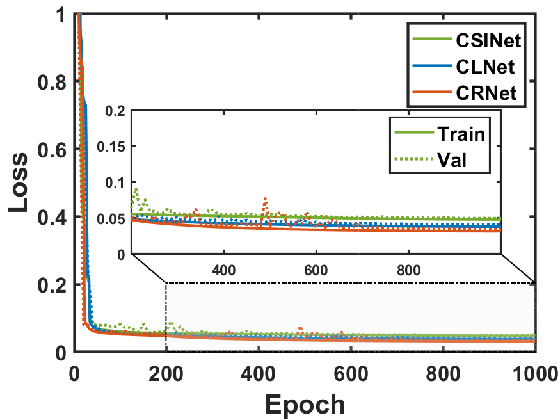}
		\centering
		\caption{Indoor}
		\label{fig:in}
	\end{subfigure}
	\begin{subfigure}[b]{0.49\linewidth}
		\includegraphics[width=\linewidth]{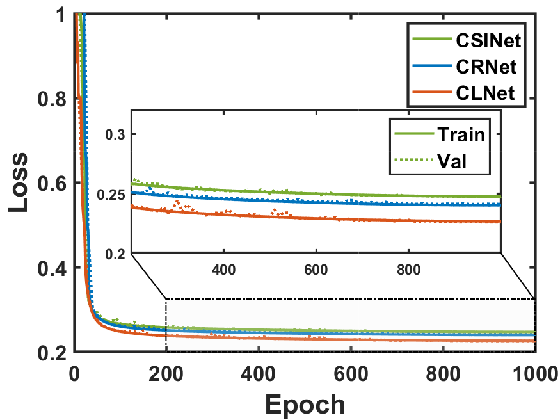}
		\centering
		
		\caption{Outdoor}
		\label{fig:out}
	\end{subfigure}
    \caption{Training loss descending trends and validation loss descending trends of three SOTA approach with JPTS adopted.}
    \label{fig:loss}
    \vspace{-0.4cm}
    
\end{figure}

\subsection{\textbf{Close Look at The Training Log}}
Because the CSI matrix in the dataset all with the dimension of 32 $\times$ 32, we choose $n=4$ tiles to get the permutation version of original CSI matrixs. The extra zero-padding operation to make the CSI matrix with the dimension of 33 $\times$ 33 is needed if $n=9$, which in fact fervor the learning process because zero is easy to learn and reconstruct like those sparse parts. As $n=9$ is a common practice in the representative learning domain~\cite{noroozi2016unsupervised} and $4! < 9!$ , we thus take a close look at the training process and want to verify if the JPTS has overfitting phenomenon. Figure~\ref{fig:loss} plots the training losses and the correspond validation losses of three SOTA approaches with JPTS adopted under compression ratio $\eta = 1/16$. Figure~\ref{fig:in} and Figure~\ref{fig:out} correspond to indoor and outdoor scenarios. we can clearly see that the JPTS works well and does not have any overfitting phenomenon and it expects to work even better if divided into 9 tiles.


\begin{figure}[t]
  \centering
  \includegraphics[width=0.9\linewidth]{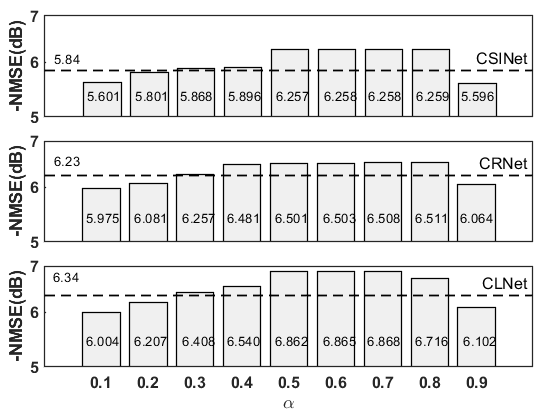}
  \caption{The NMSE(dB) of three SOTA approaches with different $\alpha$ settings. The dashed line indicates original result without adopting JPTS.}
  \label{fig:alpha}
  \vspace{-0.3cm}
\end{figure}

\subsection{\textbf{The Impact of Hyperparameter $\alpha$}}
Since we utilize $\alpha$ as a control parameter to determine the difficulty of the puzzle solving task, we perform training with different $\alpha$ to understand its impact\footnote{Note that the results in Table I all with $\alpha=0.5$.}. Fig.~\ref{fig:alpha} is the result with 1/64 compression ratio in indoor environment. The dashed line indicates the original results of three corresponding SOTA approaches. Overall, we find that the impact of $\alpha$ is consistent across three different deep learning frameworks. The proposed jigsaw puzzles aided training strategy is generally helpful when $\alpha$ is set to between 0.3 and 0.8.

There is not much difference when $\alpha$ is between 0.5 to 0.8, where JPTS-CSINet and JPTS-CRNet gain the highest performance improvement when $\alpha=0.8$, while JPTS-CLNet achieves the highest performance when $\alpha=0.7$. We see from the results that JPTS is not sensitive to the selection of $\alpha$, and the range between 0.5 to 0.8 may provide performance gains very close to the peak values. When $\alpha$ is too low, the original compression task becomes puzzle-solving task, so the goal does not align well between training and testing. On the other hand, when $\alpha$ is too high, the network has to pay attention to the extra puzzle-solving task but without enough ability to solve it.

In addition, we also evaluate the $\alpha$ at different compression ratios for a same network, JPTS-CLNet, and the results are shown in Table~\ref{t:alpha}. We observe similar results, i.e., the peak performance gain appears when $\alpha$ is set to 0.7 or 0.8, but it is not very sensitive to the specific choice of $\alpha$.

\begin{table}[t]
\centering

\small
\begin{tabular}{c|rrrrr} 
\hline\hline
\multicolumn{1}{c|}{\diagbox[innerwidth=0.9cm]{$\alpha$}{$\eta$}} & \multicolumn{1}{c}{1/4} & \multicolumn{1}{c}{1/8} & \multicolumn{1}{c}{1/16} & \multicolumn{1}{c}{1/32} & \multicolumn{1}{c}{1/64}  \\ 
\hline\hline
0.5                                         & -28.38                  & -16.03                  & -12.16                   & -9.00                    & -6.86                     \\ 
\hline
0.6                                         &      -28.68                   &    -16.07                     &      -12.22                    &          -9.00                &     -6.87                      \\ 
\hline
0.7                                         &       \textbf{-28.91}                  &        -16.13                 &             -12.22             &    -9.01                      &               \textbf{-6.87}            \\ 
\hline
0.8                                         &                   -28.82      &         \textbf{-16.18}                &                 \textbf{-12.25}         &      \textbf{-9.03}                    &             -6.72              \\
\hline
\end{tabular}
\caption{JPTS-CLNet NMSE(dB) comparison with different $\alpha$ settings at different compression ratios.}
\label{t:alpha}
\end{table}


\subsection{\textbf{The Alternative Puzzle Solving Design}}
An alternative way to solve the jigsaw puzzles is directly reconstructing the CSI from the permutation version of CSI. In such a case, the corresponding equation~\ref{e:all} turns to: 
\begin{equation}
\begin{aligned}
\footnotesize
\left(\hat{\mathit{\Theta}}_{\mathcal{E}}, \hat{\mathit{\Theta}}_{\mathcal{D}}\right)=\underset{\mathit{\Theta}_{\mathcal{E}}, \mathit{\Theta}_{\mathcal{D}}}{\arg \min }
[ \alpha \left\|\textit{\textbf{H}}_{a}-f_{\mathcal{D}}\left(f_{\mathcal{E}}\left(\textit{\textbf{H}}_{a}, \mathit{\Theta}_{\mathcal{E}}\right), \mathit{\Theta}_{\mathcal{D}}\right)\right\|_{2}^{2} \\
\footnotesize
+ (1-\alpha) \left\|\textit{\textbf{H}}_{a}-f_{\mathcal{D}}\left(f_{\mathcal{E}}\left(\overline{\textit{\textbf{{H}}}_{a}}, \mathit{\Theta}_{\mathcal{E}}\right), \mathit{\Theta}_{\mathcal{D}}\right)\right\|_{2}^{2} ]
\end{aligned}
\end{equation}
We evaluate such an alternative method as well. Table~\ref{t:alt} reports the results.

We find that either solving the puzzle directly (alternative) or solving the puzzle by identifying the index of tiles (JPTS) can assist the CSI feedback task to get performance improvement. However, the proposed JPTS can provide a more significant improvement.
\begin{table}[t]
\centering

\small
\begin{tabular}{crrr} 
\hline\hline
1/32   & \multicolumn{1}{c}{Original} & \multicolumn{1}{c}{Alternative} & \multicolumn{1}{c}{JPTS}  \\ 
\hline\hline
CSINet & -6.24                    & -8.02                    & \textbf{-8.59}                     \\ 
\hline
CRNet  & -8.90                    & -8.90                    & \textbf{-8.98}                     \\ 
\hline
CLNet  & -8.95                    & -8.96                    & \textbf{-9.00}                     \\
\hline
\end{tabular}
\caption{Indoor NMSE(dB) comparison among the original approches, when enhanced by  alternative puzzle-solving method, and enhanced by the proposed JPTS method, with 1/32 compression ratios.}
 \label{t:alt}
\end{table}

\section{Conclusion}
This paper proposes a novel jigsaw puzzles aided training strategy (JPTS) for DL-based massive MIMO CSI feedback task. The JPTS aims to promote better CSI compression by maximizing the mutual information in a local way with the fact that the relative position information matters in CSI matrix. The experiment results of adopting JPTS to different SOTA approaches demonstrate the efficacy of JPTS as well as support the argument of this paper, which is massive MIMO CSI feedback is a compression task, other than adopting powerful deep learning building blocks to better learn the most distinguishable parts of the CSI matrix, to preserve the information of entire CSI matrix as complete (but approximated) as possible also matters.

As JPTS is a training strategy that can be generally applied to any DL-based massive MIMO CSI feedback approaches, we open source the code to boost the following research as well as for reproducible.


\section*{Acknowledgment}
This work is supported by Singapore MOE AcRF Tier 2 MOE-T2EP20220-0004 and MOE AcRF Tier 1 RT13/20.


\bibliographystyle{IEEEtran}
\bibliography{bitex}




\end{document}